\documentclass[10pt,english]{iopart}
\usepackage[T1]{fontenc}
\usepackage[latin1]{inputenc}
\usepackage{graphicx}

\makeatletter

\newcommand{\boldsymbol}[1]{\mbox{\boldmath $#1$}}

 \usepackage{verbatim}

\usepackage{rotating,cite,graphicx}
\usepackage{epsfig}

\def\PRC{{\em Phys. Rev.} C}
\def\NPA{{\em Nucl. Phys.} A}
\def\AP{\em Ann. Phys. (N.Y.)}
\def\PRL{{\em Phys. Rev. Lett.}}
\def\Journal#1#2#3#4{{#1} {\bf #2}, #3 (#4)}

\def\be{\begin{equation}}
\def\ee{\end{equation}}
\def\bea{\begin{eqnarray}}
\def\eea{\end{eqnarray}}

\usepackage{babel}
\makeatother
\begin{document}

\title{A New Look at Nuclear Supersymmetry through Transfer Experiments}

\author{J. Barea$^{1)}$, R. Bijker$^{1)}$ and A. Frank$^{1,2)}$}

\address{$^{1)}$Instituto de Ciencias Nucleares, 
Universidad Nacional Autónoma de México,\\
A.P. 70-543, 04510 México, D.F.\\
$^{2)}$Centro de Ciencias Físicas, 
Universidad Nacional Autónoma de México,\\
A.P. 139-B, 62251 Cuernavaca, Morelos, México}

\begin{abstract}
Nuclear supersymmetry is reviewed and some of its
applications and extensions are discussed, together with a
proposal for new, more stringent and precise tests to probe the
supersymmetry classification, in particular, correlations between nuclei 
that belong to the same supermultiplet. The combination of these
theoretical and experimental studies may play a unifying role in
nuclear phenomena.
\end{abstract}

\maketitle

\section{Introduction}

Supersymmetric quantum mechanics (SSQM) arose from the concept
of supersymmetry in quantum field theory applied to the
simpler case of quantum mechanics \cite{ssqm}. This framework has been
very fruitful in studying potential problems in quantum mechanics,
not only to understand the connections between analytically
solvable problems, but also to discover new solutions.

In this article we discuss a somewhat different application of the
concept of supersymmetric quantum mechanics, proposed more than two 
decades ago in the field of nuclear structure physics \cite{FI}
and known as
nuclear supersymmetry (n-SUSY). This approach has similarities to
SSQM, but some significant differences too. While both frameworks
treat bosonic and fermionic systems on an equal footing, in the
traditional SSQM approach the Hamiltonian $H$ is factorized in
terms of the so called supercharges (concretely, $H$ is the 
anticommutator of the supercharges), whereas in n-SUSY the
Hamiltonian is more general and is a function of the generators of 
the graded Lie algebra associated to the supergroup which governs 
the algebraic structure of the problem.  
In analogy to the case of SSQM, in n-SUSY the fermionic generators 
of the graded Lie algebra play the role of supercharges which connect 
bosonic and fermionic systems. Physically they are associated 
with one-nucleon transfer operators connecting states in different 
neighboring nuclei. In general, however, the
supercharges do not commute with the Hamiltonian. As a
consequence, the spectra of the systems under study are not
identical, as they are in SSQM. In other words, while in SSQM $H$ is a
generator of the superalgebra, this is not the case in n-SUSY,
where a more complicated structure is used. This is a necessary
characteristic, because n-SUSY connects the spectroscopic properties 
of states in even- and odd-mass nuclei, 
and we know these properties are quite different.

In the present paper, we first describe the formalism of nuclear 
supersymmetry. Next we report the first results of an ongoing
investigation of one- and two-nucleon transfer reactions in the 
Pt-Au mass region which is considered to provide the best examples 
of n-SUSY in nature. We establish new correlations between transfer 
reactions among different pairs of nuclei as a consequence of 
n-SUSY which can be tested directly in future experiments.
Finally, we discuss future perspectives for nuclear supersymmetry, 
in particular related to some ideas put forward several years 
ago to generalize n-SUSY to other (transitional) regions of the 
nuclear mass table \cite{once}, and to special correlations between  
one- and two-nucleon transfer reactions and $\beta$ decay. 

\section{Dynamical Supersymmetries in Nuclear Physics}

Dynamical supersymmetries were introduced \cite{FI} in nuclear physics in 
1980 by Iachello in the context of the Interacting Boson Model (IBM) 
\cite{IBM} and its extensions. The spectroscopy of atomic nuclei is 
characterized by the interplay between collective (bosonic) and 
single-particle (fermionic) degrees of freedom. 

The IBM describes collective excitations in even-even nuclei in 
terms of a system of interacting monopole and quadrupole bosons with angular 
momentum $l=0,2$. The bosons are associated with the number of 
correlated valence proton and neutron pairs, 
and hence the number of bosons $N$ is 
half the number of valence nucleons. Since it is convenient to express 
the Hamiltonian and other operators of interest in second quantized form, 
we introduce creation, $s^{\dagger}$ and $d^{\dagger}_m$, and annihilation, 
$s$ and $d_m$, operators, which altogether can be denoted by 
$b^{\dagger}_{i}$ and $b_{i}$ with $i=l,m$ ($l=0,2$ and $-l \leq m \leq l$). 
The operators $b^{\dagger}_{i}$ and $b_{i}$ satisfy the commutation 
relations 
\begin{equation}
[b_i,b^{\dagger}_j] \;=\; \delta_{ij} ~, 
\hspace{1cm} [b^{\dagger}_i,b^{\dagger}_j] 
\;=\; [b_i,b_j] \;=\; 0 ~.
\end{equation}
The bilinear products 
\begin{equation}
B_{ij} \;=\; b^{\dagger}_i b_j ~,
\label{bosgen}
\end{equation} 
generate the algebra of $U(6)$ the unitary group in 6 dimensions
\begin{equation}
[ B_{ij},B_{kl} ] \;=\; B_{il} \, \delta_{jk} - B_{kj} \, \delta_{il} ~.
\label{algu6}
\end{equation}
The IBM Hamiltonian and other operators of interest are expressed 
in terms of the generators of $U(6)$. 
In general, the Hamiltonian has to be diagonalized numerically to 
obtain the energy eigenvalues and wave functions. There exist, however, 
special situations in which the eigenvalues can be obtained in closed, 
analytic form. These special solutions provide a framework in which 
energy spectra and other nuclear properties (such as quadrupole transitions 
and moments) can be interpreted in a qualitative way. 
These situations correspond to dynamical symmetries of the Hamiltonian 
\cite{IBM}. A dynamical symmetry arises, when the Hamiltonian is expressed 
in terms of Casimir invariants of a chain of subgroups of $G=U(6)$, 
$G \supset G_1 \supset G_2 \supset \ldots$ only. The eigenstates can then 
be classified uniquely according to the irreducible representations of 
$G$ and its subgroups $G_1$, $G_2$, $\ldots$. The different representations 
of $G$, $G_1$, $G_2$ $\ldots$ are split but not admixed by the 
Hamiltonian. The energy eigenvalues are given by the 
expectation values of the Casimir operators. In addition, by using standard 
group theoretical techniques it is possible to obtain analytic expressions 
for electromagnetic transition rates and quadrupole moments, etc.

The concept of dynamical symmetry has been shown to be a very useful tool 
in different branches of physics. A well-known example in nuclear physics 
is the Elliott $SU(3)$ model \cite{Elliott} to describe the properties 
of light nuclei in the $sd$ shell. Another example is the $SU(3)$ flavor 
symmetry of Gell-Mann and Ne'eman \cite{gmn} to classify the baryons 
and mesons into flavor octets, decuplets and singlets and to describe 
their masses with the Gell-Mann-Okubo mass formula. 

For odd-mass nuclei the IBM has been extended to include single-particle
degrees of freedom \cite{IBFM}. The Interacting Boson-Fermion Model (IBFM)
has as its building blocks a set of $N$ bosons with $l=0,2$ and an odd 
nucleon (either a proton or a neutron) occupuying the single-particle 
orbits with angular momenta $j=j_1,j_2,\dots$. The components of the 
fermion angular momenta span the $m$-dimensional space of the group 
$U(m)$ with $m=\sum_j (2j+1)$. 
We introduce, in addition to the boson operators for the collective 
degrees of freedom, fermion creation $a^{\dagger}_i$ and annihilation 
$a_{i}$ operators for the extra nucleon. The fermion operators satisfy 
anti-commutation relations 
\begin{equation}
\{a_i,a^{\dagger}_j\} \;=\; \delta_{ij} ~, 
\hspace{1cm} \{a^{\dagger}_i,a^{\dagger}_j\} 
\;=\; \{a_i,a_j\} \;=\; 0 ~.
\end{equation}
The bilinear products 
\begin{equation}
A_{ij} \;=\; a^{\dagger}_i a_j ~,
\label{fergen}
\end{equation} 
generate the algebra of $U(m)$, the unitary group in $m$ dimensions
\begin{equation}
[ A_{ij},A_{kl} ] \;=\; A_{il} \, \delta_{jk} - A_{kj} \, \delta_{il} ~.
\label{algum}
\end{equation}
By construction the fermion operators commute with the boson operators. 
\begin{equation}
[ B_{ij},A_{kl} ] \;=\; 0 ~.
\label{algu6um}
\end{equation}
The operators $B_{ij}$ and $A_{ij}$ generate the Lie algebra of the 
symmetry group $G=U^B(6) \otimes U^F(m)$ of the IBFM. The dynamical 
symmetries that can arise in the IBFM are known under the name of 
dynamical boson-fermion symmetries for odd-mass nuclei. 

Boson-fermion symmetries can further be extended by introducing the concept 
of supersymmetries \cite{susy}, in which states in both even-even and 
odd-even nuclei are treated in a single framework. So far, 
we have discussed the symmetry properties of a mixed system of boson and 
fermion degrees of freedom for a fixed number of bosons $N$ and one fermion 
$M=1$. The operators $B_{ij}$ and $A_{ij}$ can only change bosons into 
bosons and fermions into fermions. In addition to $B_{ij}$ and $A_{ij}$, 
one can introduce operators that change a boson into a fermion and vice 
versa, but conserve the total number of bosons and fermions 
\begin{equation}
F_{ij} \;=\; b^{\dagger}_i a_j ~, \hspace{1cm} 
G_{ij} \;=\; a^{\dagger}_i b_j ~.
\label{fgen}
\end{equation} 
The enlarged set of operators $B_{ij}$, $A_{ij}$, $F_{ij}$ and $G_{ij}$ 
forms a closed algebra which consists of both commutation and 
anticommutation relations 
\begin{eqnarray}
\, [ B_{ij}, B_{kl} ] &=& B_{il} \, \delta_{jk} - B_{kj} \, \delta_{il} ~,
\nonumber\\
\, [ B_{ij}, A_{kl} ] &=& 0 ~,
\nonumber\\
\, [ B_{ij}, F_{kl} ] &=& F_{il} \, \delta_{jk} ~,
\nonumber\\
\, [ B_{ij}, G_{kl} ] &=& -G_{kj} \, \delta_{il} ~,
\nonumber\\
\, [ A_{ij}, A_{kl} ] &=& A_{il} \, \delta_{jk} - A_{kj} \, \delta_{il} ~,
\nonumber\\
\, [ A_{ij}, F_{kl} ] &=& -F_{kj} \, \delta_{il} ~,
\nonumber\\
\, [ A_{ij}, G_{kl} ] &=& G_{il} \, \delta_{jk} ~,
\nonumber\\
\{ F_{ij}, F_{kl} \}  &=& 0 ~,
\nonumber\\
\{ F_{ij}, G_{kl} \}  &=& B_{il} \, \delta_{jk} + A_{kj} \, \delta_{il} ~,
\nonumber\\
\{ G_{ij}, G_{kl} \}  &=& 0 ~.
\label{graded}
\end{eqnarray}
This algebra can be identified with that of the graded Lie group $G=U(6/m)$. 
It provides an elegant scheme in which the IBM and IBFM can be unified 
into a single framework \cite{susy}
\begin{equation}
G=U(6/m) \supset U^B(6) \otimes U^F(m) ~. 
\end{equation}
In this supersymmetric framework, 
even-even and odd-mass nuclei form the members of a supermultiplet which 
is characterized by $[{\cal N}\}$ where ${\cal N}=N+M$, 
i.e. the total number of bosons and fermions. 
Thus, supersymmetry distinguishes itself from other symmetries 
in that it includes, in addition to transformations among fermions and among 
bosons, also transformations that change a boson into a fermion and 
vice versa. 

The Hamiltonian of n-SUSY is written in terms of the generators of 
the graded Lie algebra of $U(6/m)$ of Eq.~(\ref{graded}). 
A dynamical supersymmetry arises when the Hamiltonian is composed of 
the Casimir operators of a chain of subgroups of $U(6/m)$. 
Dynamical nuclear supersymmetries correspond to very special forms 
of the Hamiltonian which may not be applicable to all regions of the 
nuclear chart, but nevertheless several nuclei in the Os-Ir-Pt-Au region 
have been found to provide experimental evidence for the 
approximate occurrence of 
supersymmetries in nuclei. 

\section{Dynamical Neutron-Proton Supersymmetry}

\begin{figure}
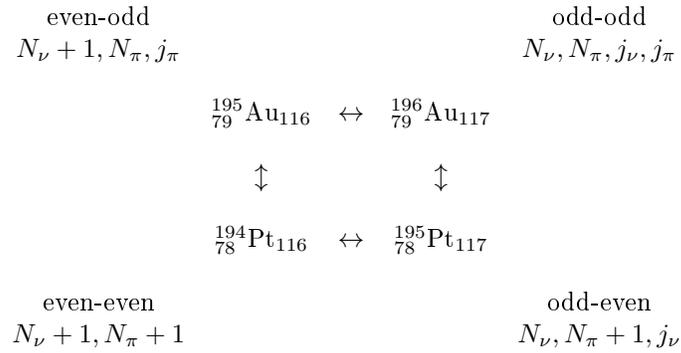

\begin{eqnarray}
\begin{array}{ccccc}
& & & & \\
\mbox{even-odd} & & & & \mbox{odd-odd} \\
N_{\nu}+1, N_{\pi}, j_{\pi} & & & & N_{\nu}, N_{\pi}, j_{\nu}, j_{\pi} \\
& & & & \\
& ^{195}_{ 79}\mbox{Au}_{116} & \leftrightarrow 
& ^{196}_{ 79}\mbox{Au}_{117} & \\
& & & & \\
& \updownarrow & & \updownarrow & \\
& & & & \\
& ^{194}_{ 78}\mbox{Pt}_{116} & \leftrightarrow 
& ^{195}_{ 78}\mbox{Pt}_{117} & \\
& & & & \\
\mbox{even-even} & & & & \mbox{odd-even} \\
N_{\nu}+1, N_{\pi}+1 & & & & N_{\nu}, N_{\pi}+1, j_{\nu} 
 \\
& & & & 
\end{array}
\nonumber
\end{eqnarray}
\caption[]{Magic quartet of nuclei}
\label{magic}
\end{figure}

The mass region $A\sim190$ has been a rich source of possible empirical
evidence for the existence of (super)symmetries in nuclei. The even-even
nucleus $^{196}$Pt is the standard example of the $SO(6)$ dynamical
symmetry (DS) of the IBM \cite{so6}. The odd-proton nuclei $^{191,193}$Ir
and $^{193,195}$Au were suggested as examples of the $Spin(6)$ DS
\cite{FI}, in which the odd-proton is allowed to occupy the $\pi d_{3/2}$
orbit, whereas the pairs of nuclei $^{192}$Os - $^{191}$Ir, $^{194}$Os
- $^{193}$Ir, $^{192}$Pt - $^{193}$Au and $^{194}$Pt - $^{195}$Au
have been analyzed as examples of a $U(6/4)$ supersymmetry \cite{susy}.
The odd-neutron nucleus $^{195}$Pt, together with $^{194}$Pt, were
studied in terms of a $U(6/12)$ supersymmetry, in which the odd neutron
occupies the $\nu p_{1/2}$, $\nu p_{3/2}$ and $\nu f_{5/2}$ orbits
\cite{baha}. These ideas were later extended to the case where neutron
and proton bosons are distinguished \cite{quartet}, predicting in
this way a correlation among quartets of nuclei, consisting of an
even-even, an odd-proton, an odd-neutron and an odd-odd nucleus. The
best experimental example of such a quartet with 
$U(6/12)_{\nu}\otimes U(6/4)_{\pi}$ supersymmetry is provided by the 
nuclei $^{194}$Pt, $^{195}$Au, $^{195}$Pt and $^{196}$Au which are  
characterized by ${\cal N}{\pi}=N_{\pi}+1=2$ and 
${\cal N}{\nu}=N_{\nu}+1=5$, see Fig.~\ref{magic}.  

The supersymmetric classification of nuclear levels in the Pt and
Au isotopes has been re-examined by taking advantage of the significant
improvements in experimental capabilities developed in the last decade.
High resolution transfer experiments with protons and polarized deuterons
have led to strong evidence for the existence of supersymmetry (SUSY)
in atomic nuclei. The experiments include high resolution transfer
experiments to $^{196}$Au at TU/LMU M\"{u}nchen \cite{metz,pt195},
and in-beam gamma ray and conversion electron spectroscopy following
the reactions $^{196}$Pt$(d,2n)$ and $^{196}$Pt$(p,n)$ at the
cyclotrons of the PSI and Bonn \cite{au196}. These studies have achieved
an improved classification of states in $^{195}$Pt and $^{196}$Au
which give further support to the original ideas \cite{baha,sun,quartet}
and extend and refine previous experimental work 
\cite{mauthofer,jolie,rotbard} in this research area.

As we mentioned before, the Pt and Au nuclei have been described in terms
of a dynamical $U(6/12)_{\nu}\otimes U(6/4)_{\pi}$ supersymmetry. 
In this case, the relevant subgroup chain is given by \cite{quartet}
\bea
U(6/12)_{\nu} \otimes U(6/4)_{\pi} 
&\supset& U^{B_{\nu}}(6) \otimes U^{F_{\nu}}(12)
\otimes
U^{B_{\pi}}(6) \otimes U^{F_{\pi}}(4) \nonumber\\
&\supset& U^B(6) \otimes U^{F_{\nu}}(6) \otimes U^{F_{\nu}}(2)
\otimes
U^{F_{\pi}}(4) \nonumber\\
&\supset& U^{BF_{\nu}}(6) \otimes U^{F_{\nu}}(2) \otimes U^{F_{\pi}}(4)
\nonumber\\
&\supset& SO^{BF_{\nu}}(6) \otimes U^{F_{\nu}}(2) \otimes SU^{F_{\pi}}(4)
\nonumber\\
&\supset& Spin(6) \otimes U^{F_{\nu}}(2) \nonumber\\
&\supset& Spin(5) \otimes U^{F_{\nu}}(2) \nonumber\\
&\supset& Spin(3) \otimes SU^{F_{\nu}}(2) \nonumber\\
&\supset& SU(2) ~. 
\label{chain}
\eea
The Hamiltonian is expressed in terms of the Casimir operators as 
\bea
H &=& \alpha \, C_{2U^{BF_{\nu}}(6)} + \beta \, C_{2SO^{BF_{\nu}}(6)}
+ \gamma \, C_{2Spin(6)} 
\nonumber\\
&& + \delta \, C_{2Spin(5)} + \epsilon \, C_{2Spin(3)} + \eta \,
C_{2SU(2)} ~. 
\eea
The corresponding eigenvalues describe simultaneously the excitation 
spectra of the quartet of nuclei in Fig.~\ref{magic} 
\begin{eqnarray}
E &=& \alpha \, \left[ N_1(N_1+5) + N_2(N_2+3) + N_3(N_3+1) \right] 
\nonumber\\
&& + \beta \, \left[ \Sigma_1(\Sigma_1+4) + \Sigma_2(\Sigma_2+2) 
+ \Sigma_3^2 \right] 
\nonumber\\
&& + \gamma \, \left[ \sigma_1(\sigma_1+4) + \sigma_2(\sigma_2+2) 
+ \sigma_3^2 \right] 
\nonumber\\
&& + \delta \, \left[ \tau_1(\tau_1+3) + \tau_2(\tau_2+1) \right] 
+ \epsilon \, J(J+1) + \eta \, L(L+1) ~.
\label{npsusy}
\end{eqnarray}
The coefficients $\alpha$, $\beta$, $\gamma$, $\delta$, $\epsilon$ and 
$\eta$ have been determined in a simultaneous fit of the excitation energies 
of the four nuclei of Fig.~\ref{magic}) \cite{au196}. 

In a dynamical supersymmetry, closed expressions can be derived for
energies, and selection rules and intensities for electromagnetic
transitions and single-particle transfer reactions. While a simultaneous
description and classification of these observables in terms of the
$U(6/12)_{\nu}\otimes U(6/4)_{\pi}$ supersymmetry has been shown
to be fulfilled to a good approximation for the quartet of nuclei
$^{194}$Pt, $^{195}$Au, $^{195}$Pt and $^{196}$Au, there are important
predictions still not fully verified by experiments. These tests involve
the transfer reaction intensities among the supersymmetric partners.
In the next section we concentrate on the latter and, in particular,
on the one-proton transfer reactions $^{194}$Pt $\rightarrow$ $^{195}$Au
and $^{195}$Pt $\rightarrow$ $^{196}$Au.

\section{One-Proton Transfer Reactions}

The single-particle transfer operator that is commonly used in the
Interacting Boson-Fermion Model (IBFM), has been derived in the seniority
scheme \cite{olaf}. Although strictly speaking this derivation is
only valid in the vibrational regime, it has been used for deformed
nuclei as well. An alternative method is based on symmetry considerations.
It consists in expressing the single-particle transfer operator in
terms of tensor operators under the subgroups that appear in the group
chain of a dynamical (super)symmetry \cite{spin6,BI,barea}. 
The use of tensor operators to describe single-particle transfer reactions 
in the supersymmetry scheme has the advantage of giving rise to selection 
rules and closed expressions for the spectroscopic factors, whose 
consequences for the experimental observables can be better gauged. 
The single-particle 
transfer between different members of the same supermultiplet provides
an important test of supersymmetries, since it involves the transformation
of a boson into a fermion or vice versa, but conserving the total
number of bosons plus fermions.

The one-proton transfer operator in the $U(6/12)_{\nu}\otimes U(6/4)_{\pi}$ 
supersymmetry consists, in lowest order, of two terms 
\bea
P^{\dagger} &=& \alpha_0 \, 
\left( \tilde{s}_{\pi} \times a^{\dagger}_{\pi,3/2} 
\right)^{(3/2)} + \alpha_2 \, \left( \tilde{d}_{\pi} 
\times a^{\dagger}_{\pi,3/2} \right)^{(3/2)} ~, 
\label{transfer}
\eea 
that  describe the one-proton transfer reactions between the Pt and Au 
nuclei belonging to the quartet of nuclei of Eq.~(\ref{magic}): 
$^{194}$Pt $\rightarrow$ $^{195}$Au and $^{195}$Pt $\rightarrow$ $^{196}$Au. 
In Table~\ref{int1} we present the intensities of the allowed one-proton 
transfer reactions from the ground state $|(N+2,0,0),(0,0),0\rangle$ of the 
even-even nucleus $^{194}$Pt to the even-odd nucleus $^{195}$Au belonging 
to the same supermultiplet $[{\cal N}_{\nu}\} \otimes [{\cal N}_{\pi}\}
=[N_{\nu}+1\} \otimes [N_{\pi}+1\}$. 
The intensity is defined as 
\bea
I \;=\;  
\left| \langle f \mid\mid P^{\dagger} \mid\mid i \rangle \right|^2 ~. 
\eea
The transfer operator of Eq.~(\ref{transfer}) is a tensor operator under 
$Spin(5)$ and $Spin(3)$. Its transformation properties are 
$(\tau_{1},\tau_{2})=(1/2,1/2)$ under $Spin(5)$ and $J=3/2$ under 
$Spin(3)$. Due to the selection rules, $P^{\dagger}$ can only excite 
states in the final nucleus with $(\tau_{1},\tau_{2})=(1/2,1/2)$ and $J=3/2$. 
The allowed values of the $Spin(6)$ labels are 
$(\sigma_{1},\sigma_{2},\sigma_{3})=(N+3/2,1/2,1/2)$ for the ground state 
and $(N+1/2,1/2,-1/2)$ for an excited state. The ratio of the intensities 
for the excitation of the excited and the ground state is given by 
\bea
R \;=\; \frac{I_{\rm gs \rightarrow exc}}{I_{\rm gs \rightarrow gs}} 
\;=\; (N+1)(N+5) \left[ \frac{\alpha_0 \sqrt{5}+\alpha_2}
{\alpha_0(N+5)\sqrt{5}-\alpha_2(N+1)} \right]^2 ~.
\label{ratio1}
\eea
The number of bosons $N$ is taken to be the number of bosons in the odd-odd 
nucleus $^{196}$Au: $N=N_{\nu}+N_{\pi}=4+1=5$, see Fig.~\ref{magic}.
In the bottom half of Table~\ref{int1}, we show the allowed transitions 
for the one-proton transfer from the ground state 
$|(N+2,0,0),(0,0),0,1/2\rangle$ of the odd-even nucleus $^{195}$Pt 
to the odd-odd nucleus $^{196}$Au. In this case, the transfer operator 
of Eq.~(\ref{transfer}) excites doublets of $^{196}$Au characterized by 
$(\tau_{1},\tau_{2})=(1/2,1/2)$, $J=3/2$ and $L=J \pm 1/2$,  
belonging to the ground band with 
$(\sigma_{1},\sigma_{2},\sigma_{3})=(N+3/2,1/2,1/2)$, 
and to an excited band with $(N+1/2,1/2,-1/2)$. The ratio 
of the intensities is the same as that for the $^{194}$Pt $\rightarrow$
$^{195}$Au transfer reaction in Eq.~(\ref{ratio1})
\bea
R(^{195}\mbox{Pt} \rightarrow ^{196}\mbox{Au}) 
\;=\; R(^{194}\mbox{Pt} \rightarrow ^{195}\mbox{Au}) ~.
\label{ratio2}
\eea 
This is a direct consequence of the supersymmetry 
classification of the states. 

\begin{figure}
\centering
\setlength{\unitlength}{1.0pt}
\begin{picture}(300,160)(0,0)
\thicklines
\put ( 50, 60) {\line(1,0){60}}
\put (200, 60) {\line(1,0){60}}
\put (200,140) {\line(1,0){60}}
\put (125,120) {$0/112$}
\put (145, 40) {$100/100$}
\put ( 60, 40) {$(7,0,0)$}
\put ( 60, 20) {$^{194}$Pt}
\put ( 10, 57) {$(0,0),0$}
\put (210, 40) {$(\frac{15}{2},\frac{1}{2},\frac{1}{2})$}
\put (210, 20) {$^{195}$Au}
\put (265, 57) {$(\frac{1}{2},\frac{1}{2}),\frac{3}{2}$}
\put (210,120) {$(\frac{13}{2},\frac{1}{2},-\frac{1}{2})$}
\put (265,137) {$(\frac{1}{2},\frac{1}{2}),\frac{3}{2}$}
\thinlines
\put (115, 60) {\vector( 1, 0){80}}
\put (115, 60) {\vector( 1, 1){80}}
\end{picture}
\caption[]{\small Allowed one-proton transfer reactions for 
$^{194}$Pt $\rightarrow$ $^{195}$Au. The spectroscopic factors 
are normalized to 100 for the ground state to ground state 
transition for the operators $P_1/P_2$.}
\label{spec1}
\end{figure}
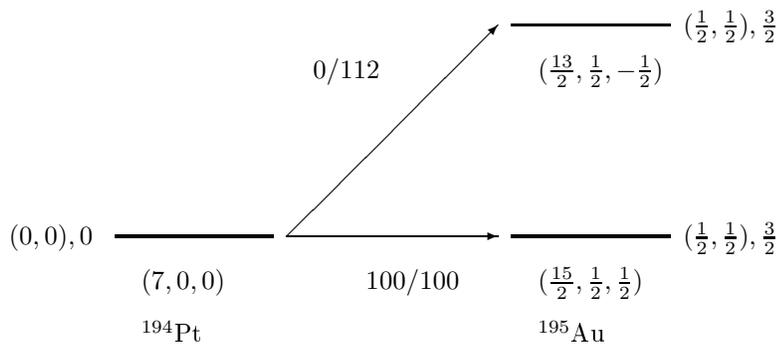

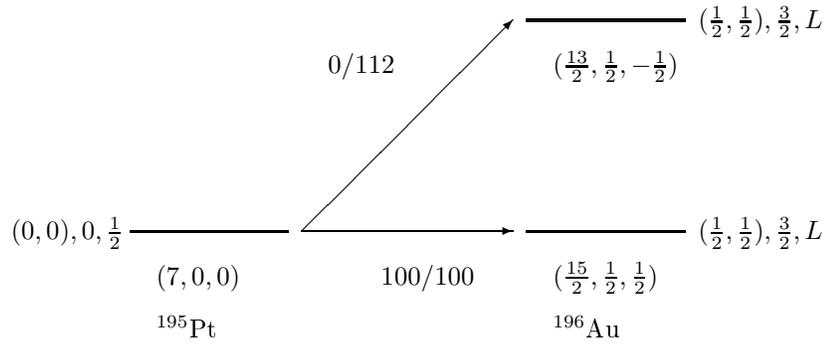
\begin{figure}[t]
\centering
\setlength{\unitlength}{1.0pt}
\begin{picture}(300,160)(0,0)
\thicklines
\put ( 50, 60) {\line(1,0){60}}
\put (200, 60) {\line(1,0){60}}
\put (200,140) {\line(1,0){60}}
\put (125,120) {$0/112$}
\put (145, 40) {$100/100$}
\put ( 60, 40) {$(7,0,0)$}
\put ( 60, 20) {$^{195}$Pt}
\put (  5, 57) {$(0,0),0,\frac{1}{2}$}
\put (210, 40) {$(\frac{15}{2},\frac{1}{2},\frac{1}{2})$}
\put (210, 20) {$^{196}$Au}
\put (265, 57) {$(\frac{1}{2},\frac{1}{2}),\frac{3}{2},L$}
\put (210,120) {$(\frac{13}{2},\frac{1}{2},-\frac{1}{2})$}
\put (265,137) {$(\frac{1}{2},\frac{1}{2}),\frac{3}{2},L$}
\thinlines
\put (115, 60) {\vector( 1, 0){80}}
\put (115, 60) {\vector( 1, 1){80}}
\end{picture}
\caption[]{\small As Fig.~\ref{spec1}, but for 
$^{195}$Pt $\rightarrow$ $^{196}$Au.}
\label{spec2}
\end{figure}

For special choices of $\alpha_0$ and $\alpha_2$, the transfer operator 
of Eq.~(\ref{transfer}) becomes a tensor operator under $Spin(6)$ as well
\bea
P^{\dagger}_{1} &=& \alpha \left[ -\sqrt{\frac{1}{6}} 
\left( \tilde{s}_{\pi} \times a^{\dagger}_{\pi,3/2} 
\right)^{(3/2)} +\sqrt{\frac{5}{6}} \left( \tilde{d}_{\pi} 
\times a^{\dagger}_{\pi,3/2} \right)^{(3/2)} \right] ~, 
\nonumber\\ 
P^{\dagger}_{2} &=& \alpha \left[ +\sqrt{\frac{5}{6}} 
\left( \tilde{s}_{\pi} \times a^{\dagger}_{\pi,3/2} 
\right)^{(3/2)} +\sqrt{\frac{1}{6}} \left( \tilde{d}_{\pi} 
\times a^{\dagger}_{\pi,3/2} \right)^{(3/2)} \right] ~.  
\label{tensor} 
\eea
Here $P^{\dagger}_{1}$ transforms as 
$(\sigma_{1},\sigma_{2},\sigma_{3})=(1/2,1/2,-1/2)$ under $Spin(6)$, and 
$P^{\dagger}_{2}$ as $(3/2,1/2,1/2)$. Due to the $Spin(6)$ selection rules, 
the operator $P^{\dagger}_{1}$ only excites the ground state 
of the Au nuclei with $(\sigma_1,\sigma_2,\sigma_3)=(N+3/2,1/2,1/2)$, 
whereas $P^{\dagger}_{2}$ populates, in addition to the ground state, 
also an excited state with $(N+1/2,1/2,-1/2)$. 
In Table~\ref{int2}, we present the intensities of the allowed 
transfers for the operators of Eq.~(\ref{tensor}). These correspond 
to special cases of the more general results of Table~\ref{int1}. 
Figs.~\ref{spec1} and~\ref{spec2} show the allowed transitions 
for the one-proton transfer reaction $^{194}$Pt $\rightarrow$ $^{195}$Au 
and $^{195}$Pt $\rightarrow$ $^{196}$Au, respectively. 
The ratio of the intensities is now given by  
\bea
R_1(^{195}\mbox{Pt} \rightarrow ^{196}\mbox{Au}) 
= R_1(^{194}\mbox{Pt} \rightarrow ^{195}\mbox{Au}) 
\;=\; 0 ~, 
\nonumber\\
R_2(^{195}\mbox{Pt} \rightarrow ^{196}\mbox{Au}) 
= R_2(^{194}\mbox{Pt} \rightarrow ^{195}\mbox{Au}) 
\;=\; \frac{9(N+1)(N+5)}{4(N+6)^2} ~, 
\label{ratios} 
\eea
for $P_{1}$ and $P_{2}$, respectively. For the one-proton transfer reactions 
$^{194}$Pt $\rightarrow$ $^{195}$Au and $^{195}$Pt $\rightarrow$ $^{196}$Au, 
the second ratio is given by $R_{2}=1.12$ ($N=5$). 

The available experimental data from the proton stripping
reactions $^{194}$Pt$(\alpha,t)^{195}$Au and
$^{194}$Pt$(^{3}$He$,d)^{195}$Au \cite{munger} shows that the
$J=3/2$ ground state of $^{195}$Au is excited strongly with
$C^{2}S=0.175$, whereas the first excited $J=3/2$ state is excited
weakly with $C^{2}S=0.019$. In the SUSY scheme, the latter state
is assigned as a member of the ground state band with
$(\tau_{1},\tau_{2})=(5/2,1/2)$. Therefore the one proton transfer
to this state is forbidden by the $Spin(5)$ selection rule of the
tensor operators of Eq.~(\ref{tensor}). The relatively small
strength to excited $J=3/2$ states suggests that the operator
$P_{1}$ of Eq.~(\ref{tensor}) can be used to describe the data with
a good degree of approximation. 

According to Eq.~(\ref{ratio2}) and~(\ref{ratios}), the ratio of the 
intensities for the $^{195}$Pt $\rightarrow$ $^{196}$Au transfer reaction 
is the same as that for $^{194}$Pt $\rightarrow$ $^{195}$Au. The equality 
of the ratios is a consequence of the supersymmetry classification. This 
prediction will be tested experimentally using the $(^3$He$,d)$ 
reaction on $^{194}$Pt and $^{195}$Pt targets \cite{graw}. 

\section{Correlations}

As we have seen in the previous section, the matrix elements for one-proton 
transfer reactions between odd-neutron and odd-odd nuclei are related to 
those between and even-even and odd-proton nuclei. The results were 
obtained by deriving the matrix elements and taking the ratios. However, 
it is possible to generalize these results and to establish explicit 
relations between the intensities of these two transfer reactions, 
i.e. the one-proton transfer reaction intensities between the (ground state 
of the) Pt and Au nuclei are related by 
\bea
I(^{195}\mbox{Pt} \, \rightarrow \, ^{196}\mbox{Au}) &=& \frac{2L+1}{4} \, 
I(^{194}\mbox{Pt} \, \rightarrow \, ^{195}\mbox{Au}) ~.
\label{corr}
\eea
This correlation holds for both the general form of the transfer operator 
of Eq.~(\ref{transfer}) and the two tensor operators of Eq.~(\ref{tensor}). 
It can be derived from the symmetry relations that exist between the 
different $U(6)$ couplings in the wave functions of the even-even, odd-even, 
even-odd and odd-odd nuclei of a supersymmetric quartet (the so-called 
$F$-spin properties \cite{IBM}). As a consequence, it is sufficient to 
derive the intensities for one of the reactions only. The intensities for 
the other reaction can then be obtained immediately from the correlation 
in Eq.~(\ref{corr}). 
 
For the one-neutron transfer reactions, $^{194}$Pt $\leftrightarrow$ 
$^{195}$Pt and $^{195}$Au $\leftrightarrow$ $^{196}$Au, there exists a 
similar situation. We have found some preliminary results for correlations 
among different reactions which are similar, but not identical, to those 
obtained for one-proton transfer in Eq.~(\ref{corr}).  

There are still two other possible tests that probe directly the 
structure of the wave functions of a supermultiplet of nuclei: 
(i) the two-nucleon transfer reaction $^{194}$Pt$(\alpha,d)^{196}$Au 
that has been measured recently \cite{graw}, in which a
neutron-proton pair is transferred to the target nucleus. This
reaction presents a very sensitive test of the wave functions, 
since it is not only a measure for the transfer intensity, but it   
also probes the correlation within the transferred neutron-proton pair. 
(ii) the charge-exchange reaction $^{195}$Au $\rightarrow$ $^{195}$Pt 
(also connected to the $\beta$ decay which has 
been studied in the IBFM in \cite{dobes}). Theoretically, both processes 
involve a combination of the operator for one-proton and for one-neutron 
transfer reactions inside the same supermultiplet. 

In principle, the available experimental data from the proton stripping
reactions $^{194}$Pt$(\alpha,t)^{195}$Au and 
$^{194}$Pt$(^{3}$He$,d)^{195}$Au \cite{munger} and from the neutron 
stripping reaction $^{194}$Pt$(d,p)^{195}$Pt 
\cite{sheline} can be used to determine the appropriate form of the 
one-proton and one-neutron transfer operators \cite{BI}, which then can be 
used to predict the spectroscopic factors for the other one-nucleon transfer 
reactions between nuclei belonging to the quartet of Fig.~\ref{magic}, e.g. 
$^{195}$Au $\rightarrow$ $^{196}$Au and $^{195}$Pt $\rightarrow$ $^{196}$Au, 
as well as for the two-nucleon transfer reaction $^{194}$Pt$(\alpha,d)^{196}$Au 
and the $\log ft$ values of the $beta$ decay $^{195}$Au $\rightarrow$ $^{195}$Pt 
\cite{icn}. 

\section{Summary, Conclusions and Outlook}

The recent measurements of the spectroscopic properties of the odd-odd 
nucleus $^{196}$Au have rekindled the interest in nuclear supersymmetry. 
The available data on the spectroscopy of the quartet of nuclei 
$^{194}$Pt, $^{195}$Au, $^{195}$Pt and $^{196}$Au can, to a good 
approximation, be described in terms of the 
$U(6/4)_{\pi}\otimes U(6/12)_{\nu}$ supersymmetry. 
However, there is a still an important set of experiments which can 
further test the predictions of the supersymmetry scheme: 
transfer reactions between nuclei belonging to the same supermultiplet, 
in particular between the odd-even (and even-odd) and odd-odd members of 
the supersymmetric quartet. Theoretically, these transfers are described 
by the supersymmetric generators which change a boson into a fermion, or 
vice versa. Most available data involve transfer reactions 
between nuclei belonging to different multiplets. 

In this article, we investigated one-proton transfer reactions 
between the SUSY partners: $^{194}$Pt $\rightarrow$ $^{195}$Au and
$^{195}$Pt $\rightarrow$ $^{196}$Au. The supersymmetry implies
an explicit correlation between the spectroscopic factors of these two
reactions which can be tested experimentally. Preliminary results suggest 
that for the one-neutron transfer reactions $^{194}$Pt $\leftrightarrow$ 
$^{195}$Pt and $^{195}$Au $\leftrightarrow$ $^{196}$Au there exist  
correlations similar to those obtained for one-proton transfer in 
Eq.~(\ref{corr}). 
To the best of our knowledge, this is the first time that such 
relations are predicted for nuclear reactions among different pairs 
of nuclei, which may provide a
challenge and motivation for future experiments.

An extension of these ideas can be applied 
is to the two-nucleon transfer reaction $^{194}$Pt$(\alpha,d)^{196}$Au 
and the charge-exchange reaction (or $\beta$ decay) 
$^{195}$Au $\rightarrow$ $^{195}$Pt. Even though 
they may represent different physical processes, i.e. one- and two-nucleon 
transfer reactions and $\beta$ decay, the nuclear structure 
contributions are related by supersymmetry. Whether it is possible 
to find a simultaneous description in which all of these processes  
are correlated by SUSY is an open question \cite{icn}.
 
In this paper, we have discussed n-SUSY in combination with dynamical 
symmetries. However, dynamical symmetries are very scarce and have 
severely limited the study of nuclear supersymmetry. An example of 
n-SUSY without dynamical symmetry is a study of the Ru and Rh isotopes 
in the $U(6/12)$ supersymmetry, in which a combination of the 
$U^{BF}(5)$ and $SO^{BF}(6)$ dynamical symmetries was 
shown to give an excellent description of the data \cite{once}. 
This opens up the possibility to generalize n-SUSY to transitional 
regions of the nuclear mass table, to find other examples of 
supersymmetric quartets of nuclei, and to further extend the search 
for correlations as a result SUSY.  
Of course, it remains to be seen whether the correlations
predicted by n-SUSY are verified by future experiments and
whether these correlations can be truly extended to other 
regions of the nuclear table. If this is indeed the case, nuclear
supersymmetry may yet provide a powerful unifying scheme for
atomic nuclei.

\section*{Acknowledgments}

We are grateful to C. Alonso, J. Arias, G. Graw, J. Jolie, 
P. Van Isacker, and H.-F. Wirth for many discussions. 
This paper was supported in part by Conacyt, Mexico.

\clearpage

\clearpage

\begin{table}
\caption[]{Intensities of one-proton transfer reactions for the transfer
operator of Eq.~(\protect\ref{transfer}). For the supersymmetric quartet of 
nuclei $^{194,195}$Pt-$^{195,196}$Au the boson numbers are $N_{\nu}=4$, 
$N_{\pi}=1$ and $N=N_{\nu}+N_{\pi}=5$, see Fig.~\protect\ref{magic}.}
\label{int1}
\vspace{15pt}
\begin{tabular}{lc}
\hline
& \\
$^{194}$Pt $\rightarrow$ $^{195}$Au 
& $\mid \langle f || \, P^{\dagger} \, || i \rangle \mid^2$ \\
& \\
\hline
& \\
$\langle (N+\frac{3}{2},\frac{1}{2},\frac{1}{2}),
(\frac{1}{2},\frac{1}{2}),\frac{3}{2} \mid$ 
& $[\alpha_0(N+5)\sqrt{5}-\alpha_2(N+1)]^2 \, \frac{N_{\pi}+1}{5(N+3)^2}$ \\
& \\
$\langle (N+\frac{1}{2},\frac{1}{2},-\frac{1}{2}),
(\frac{1}{2},\frac{1}{2}),\frac{3}{2} \mid$ 
& $[\alpha_0 \sqrt{5}+\alpha_2]^2 \, \frac{(N+1)(N+5)(N_{\pi}+1)}{5(N+3)^2}$ \\
& \\
\hline
& \\
$^{195}$Pt $\rightarrow$ $^{196}$Au 
& $\mid \langle f || \, P^{\dagger} \, || i \rangle \mid^2$ \\
& \\
\hline
& \\
$\langle (N+\frac{3}{2},\frac{1}{2},\frac{1}{2}),
(\frac{1}{2},\frac{1}{2}),\frac{3}{2},L \mid$ 
& $[\alpha_0(N+5)\sqrt{5}-\alpha_2(N+1)]^2 \, \frac{N_{\pi}+1}{5(N+3)^2}
\frac{2L+1}{4}$ \\
& \\
$\langle (N+\frac{1}{2},\frac{1}{2},-\frac{1}{2}),
(\frac{1}{2},\frac{1}{2}),\frac{3}{2},L \mid$ 
& $[\alpha_0 \sqrt{5}+\alpha_2]^2 \, \frac{(N+1)(N+5)(N_{\pi}+1)}{5(N+3)^2} 
\frac{2L+1}{4}$ \\
& \\
\hline
\end{tabular}
\end{table}

\begin{table}
\caption[]{As Table~\protect\ref{int1}, but for the transfer operators 
of Eq.~(\protect\ref{tensor}).}
\label{int2}
\vspace{15pt}
\begin{tabular}{lcc}
\hline
& & \\
$^{194}$Pt $\rightarrow$ $^{195}$Au 
& $\mid \langle f || \, P^{\dagger}_{1} 
\, || i \rangle \mid^2$ 
& $\mid \langle f || \, P^{\dagger}_{2} 
\, || i \rangle \mid^2$ \\
& & \\
\hline
& & \\
$\langle (N+\frac{3}{2},\frac{1}{2},\frac{1}{2}),
(\frac{1}{2},\frac{1}{2}),\frac{3}{2} \mid$ 
& $\frac{2(N_{\pi}+1)}{3} \, \alpha^2$  
& $\frac{8(N+6)^2(N_{\pi}+1)}{15(N+3)^2} \, \alpha^2$ \\
& & \\
$\langle (N+\frac{1}{2},\frac{1}{2},-\frac{1}{2}),
(\frac{1}{2},\frac{1}{2}),\frac{3}{2} \mid$ & 0 
& $\frac{6(N+1)(N+5)(N_{\pi}+1)}{5(N+3)^2} \, \alpha^2$ \\
& & \\
\hline
& & \\
$^{195}$Pt $\rightarrow$ $^{196}$Au 
& $\mid \langle f || \, P^{\dagger}_{1} \, || i \rangle \mid^2$ 
& $\mid \langle f || \, P^{\dagger}_{2} \, || i \rangle \mid^2$ \\
& & \\
\hline
& & \\
$\langle (N+\frac{3}{2},\frac{1}{2},\frac{1}{2}),
(\frac{1}{2},\frac{1}{2}),\frac{3}{2},L \mid$ 
& $\frac{2(N_{\pi}+1)}{3} \frac{2L+1}{4} \, \alpha^2$ 
& $\frac{8(N+6)^2(N_{\pi}+1)}{15(N+3)^2} \frac{2L+1}{4} \, \alpha^2$ \\
& & \\
$\langle (N+\frac{1}{2},\frac{1}{2},-\frac{1}{2}),
(\frac{1}{2},\frac{1}{2}),\frac{3}{2},L \mid$ & 0 
& $\frac{6(N+1)(N+5)(N_{\pi}+1)}{5(N+3)^2} \frac{2L+1}{4} \, \alpha^2$ \\
& & \\
\hline
\end{tabular}
\end{table}

\end{document}